\documentclass[twocolumn,preprintnumbers,amsmath,amssymb,aps,prd,nofootinbib,superscriptaddress,eqsecnum,floatfix]{revtex4}
\usepackage{graphicx,color}

\makeatletter
\renewcommand{\p@subsection}{}
\makeatother

\newcommand{\Slash}[1]{\ooalign{\hfil/\hfil\crcr$#1$}}

\begin{document}

\title{Quark-Flavour Dependence of the Shear Viscosity in a Quasiparticle Model}

\author{Valeriya Mykhaylova}
\affiliation{%
Institute of Theoretical Physics, University of Wroclaw,PL-50204 Wroclaw,Poland
}
\author{Marcus Bluhm}
\affiliation{%
Institute of Theoretical Physics, University of Wroclaw,PL-50204 Wroclaw,Poland
}
\affiliation{%
SUBATECH UMR 6457 (IMT Atlantique, Universit\'e de Nantes, \\IN2P3/CNRS), 4 rue Alfred Kastler, 44307 Nantes, France
}
\author{Krzysztof Redlich}
\affiliation{%
Institute of Theoretical Physics, University of Wroclaw,PL-50204 Wroclaw,Poland
}
\author{Chihiro Sasaki}
\affiliation{%
Institute of Theoretical Physics, University of Wroclaw,PL-50204 Wroclaw,Poland
}

\date{\today}

\begin{abstract}
We study the temperature dependence of the shear viscosity to entropy density ratio in pure Yang-Mills theory and in QCD with light and strange quarks within kinetic theory in the relaxation time approximation. As effective degrees of freedom in a deconfined phase we consider quasiparticle excitations with quark and gluon quantum numbers and dispersion relations that depend explicitly on the temperature. The quasiparticle relaxation times are obtained by computing the microscopic two-body scattering amplitudes for the elementary scatterings among the quasiparticles. For pure Yang-Mills theory we show that the shear viscosity to entropy density ratio exhibits a characteristic non-monotonicity with a minimum at the first-order phase transition. In the presence of dynamical quarks the ratio smoothens while still exhibiting a minimum near confinement. Furthermore, there is a significant increase of the shear viscosity to entropy density ratio in QCD resulting from the quark contributions. This observation differs from previously reported estimates based on functional methods but is in line with perturbative QCD expectations at higher temperatures.
\end{abstract}

\maketitle

\section{Introduction\label{Sec:Intro}}

The wealth of collected experimental data in combination with first-principle results from lattice QCD evidence the creation of a strongly coupled quantum fluid, the Quark-Gluon Plasma (QGP), in the ultra-relativistic heavy-ion collisions at the Large Hadron Collider (LHC) and the Relativistic Heavy Ion Collider (RHIC)~\cite{Shuryak:2008eq,Jacak:2012dx,Andronic:2017pug}. One major goal of these experiments is to reveal the equilibrium and transport properties of the QGP as deconfined state of strongly interacting matter. In particular, the transport coefficients are sensitive to the relevant degrees of freedom, their properties and interactions in the plasma. The shear viscosity $\eta$ as a measure of the resistance against momentum modifications in the fluid represents a prominent example. Its knowledge and that of its ratio $\eta/s$ with the entropy density $s$ are important for fluid dynamical simulations. In fact, the success of applying fluid dynamics for the description of the expanding fireball created in a heavy-ion collision suggests that the QGP is in approximate local equilibrium.

Early applications of fluid dynamics confronting, in particular, elliptic flow data revealed that the QGP constitutes a nearly perfect fluid~\cite{Romatschke:2007mq,Luzum:2008cw,Heinz:2008tv,Song:2008hj,Schenke:2011tv}. The specific shear viscosity $\eta/s$ was estimated to be close to the Kovtun-Son-Starinets (KSS) lower bound~\cite{Kovtun:2004de} of $1/4\pi$ predicted by applying the duality between strongly coupled gauge and weakly coupled gravity theories. Simulations with an evolution-averaged $\eta/s$ based on comparisons with combined experimental data from top-RHIC and LHC energies \cite{Song:2008hj,Song:2011hk,Heinz:2013th} extracted a possible range of ${1<(\eta/s)/(1/4\pi)<5}$. Those estimates suffered from sizable systematic and statistical errors, cf.~\cite{Jaiswal:2016hex} for a review. More realistic studies then considered a temperature \mbox{$T$-dependence} of the transport coefficient~\cite{Niemi:2011ix,Song:2011qa}. It was found that the combined data favour an increase with $T$ up to a factor of $5$ from RHIC to LHC~\cite{Niemi:2015qia}. A possible baryo-chemical potential $\mu_B$-dependence was investigated in~\cite{Karpenko:2015xea}, finding a moderate increase with increasing $\mu_B$. The wealth of accumulated experimental data also made Bayesian estimate studies for the temperature~\cite{Bernhard:2016tnd} and chemical potential~\cite{Auvinen:2017fjw} dependence possible, confirming the previous results.

First-principle calculations of the specific shear viscosity in QCD are rather scarce. Lattice gauge theory determinations of $\eta/s$ are available only in pure Yang-Mills theory and only for a few values of $T$~\cite{Nakamura:2004sy,Meyer:2007ic,Mages:2015rea,Astrakhantsev:2017nrs}. For QCD, including dynamical quarks, no explicit calculations exist. Estimates based on the results for Yang-Mills theory and information from perturbative QCD~\cite{Arnold:2003zc} suggest a slight increase in the presence of dynamical quarks~\cite{Astrakhantsev:2017nrs,Meyer:2009jp}. As an alternative, functional diagrammatic approaches to QCD were recently exploited~\cite{Haas:2013hpa,Christiansen:2014ypa} to determine the shear viscosity in Yang-Mills theory via the Kubo relation~\cite{Kubo:1957mj} from gluon spectral functions. Those are in favour of a quasiparticle structure. The results presented in~\cite{Christiansen:2014ypa} are in reasonable agreement with the lattice results and provide an estimate for QCD with $N_f=2+1$ quark flavours also indicating only a slight increase. Both first-principle approaches find a minimal $\eta/s$ of about $0.2$ near the deconfinement transition temperature $T_c$ with a moderate increase with increasing $T$ that is qualitatively in line with the estimates from fluid dynamical simulations. We note that similar results can be obtained from perturbation theory with appropriately chosen scales in the running coupling~\cite{Christiansen:2014ypa,Jackson:2017sqw,Jackson:2017hfz}.

Since first estimates indicated that $\eta/s$ of deconfined strongly interacting matter is close to the KSS bound, various QCD-like and phenomenological approaches were studied to give an explanation for the apparent perfectness of the QGP in terms of relevant degrees of freedom in the plasma. The specific shear viscosity of quark matter was investigated in Nambu--Jona-Lasinio (NJL) models~\cite{Zhuang:1995uf,Sasaki:2008um,Marty:2013ita,Ghosh:2013cba,Ghosh:2014vja,Ghosh:2015mda,Deb:2016myz,Harutyunyan:2017ttz}. Further investigations in terms of a Gribov-Zwanziger plasma~\cite{Florkowski:2015dmm}, the Polyakov-loop improved linear sigma model \cite{Tawfik:2016edq} or a Polyakov-loop extended Quark-Meson model~\cite{Abhishek:2017pkp} were made. Moreover, kinetic theory within partonic transport simulations was exploited~\mbox{\cite{Plumari:2012ep,Plumari:2019gwq,Xu:2007ns,Xu:2007jv}}~as well as anisotropic fluid dynamics~\cite{Alqahtani:2017jwl,Alqahtani:2017mhy}, both supporting the idea of a medium composed of quasiparticle excitations.

It is a widespread paradigm that a quasiparticle description cannot account for the perfect fluidity observed for deconfined strongly interacting matter. The first quantitative determination of the specific shear viscosity for pure Yang-Mills theory described with massive quasiparticles found, refuting this paradigm, an ${\eta/s\simeq0.2}$ with a negligible $T$-dependence by using the Kubo formalism~\cite{Peshier:2005pp}. Based on the early works in~\cite{Hosoya:1983xm,Gavin:1985ph}, kinetic theory calculations in relaxation time approximation followed considering a medium composed of quasiparticles without residual mean field interaction~\cite{Sasaki:2008um,Sasaki:2008fg} and for pure Yang-Mills theory with mean field interaction term~\cite{Bluhm:2009ef,Khvorostukhin:2010cw,Bluhm:2010qf,Khvorostukhin:2011mt}. This idea was extended to describe interacting hadronic matter at vanishing~\cite{Chakraborty:2010fr} and finite chemical potential~\cite{Albright:2015fpa}. Further quasiparticle model (QPM) predictions for QCD matter were presented in~\cite{Plumari:2011mk} and, taking the possible formation of turbulences in an expanding QGP into account, in~\cite{Chandra:2012qq,Mitra:2017sjo}. Modeling quasiparticle interactions, the QPM was moreover extended by including a finite (and even large) collisional width~$\Gamma$ in the quasiparticle spectral functions. With this approach~\cite{Ozvenchuk:2012kh,Berrehrah:2016vzw,Moreau:2019vhw}, using the Kubo formalism or kinetic theory with relaxation times $\tau=1/\Gamma$, an $\eta/s$ similar to the first-principle and fluid dynamical simulation results was obtained.

In the present work, we study the temperature dependence of the specific shear viscosity of deconfined strongly interacting matter for pure Yang-Mills theory and for matter anti-matter symmetric QCD with $N_f=2+1$ quark flavours. Both systems are described in a framework with quasiparticle degrees of freedom. The shear viscosity is calculated from kinetic theory in the relaxation time approximation. The underlying quasiparticle model is outlined in Sec.~\ref{Sec:QPM}. The relaxation times are, similar to~\cite{Berrehrah:2013mua,Moreau:2019vhw}, obtained from evaluating explicitly microscopic scattering amplitudes of elementary scatterings among the quasiparticles with $T$-dependent properties and can be found in Sec.~\ref{Sec:RTA}. Our results for $\eta/s$ are presented in Sec.~\ref{Sec:Results}, where we discuss the role quark degrees of freedom play for the shear viscosity of the QGP. We summarize our findings in Sec.~\ref{Sec:Conclusions}.

\section{Quasiparticle Model\label{Sec:QPM}}

In the following we will utilize the basic version of the successful quasiparticle model~\cite{Bluhm:2004xn,Bluhm:2006yh,Bluhm:2007nu} in which equilibrium thermodynamic quantities are defined as standard phase space integrals over the thermal distribution functions of quarks and gluons which obey medium dependent dispersion relations. The thermodynamic integrals are dominated by excitations with thermal momenta $k\sim T$. Within the model, deconfined QCD matter is described by quasiparticles with effective masses and a residual mean field interaction which depend on the temperature. Longitudinal plasmon and quark-hole excitations are, instead, assumed to be exponentially suppressed~\cite{leBellac}.

Describing deconfined QCD matter with $N_f=2+1$ quark flavours at $\mu_B=0$, the entropy density in the model is given by the sum of contributions from gluons $g$, light quarks $l$ and strange quarks $s$ including their anti-particles as
\begin{equation}
 \label{e:entr0}
  s = \sum_{i = g,l,\bar{l},s,\bar{s}} s_i \,, \quad s_i = \frac{d_i}{2\pi^2} \int_0^\infty
      \!\!\! k^2 dk \frac{\left( \frac{4}{3}k^2{+}m_i^2 \right)}{E_i T}
      f_i^0 \,,
\end{equation}
where $d_i$ are the spin-colour degeneracy factors, ${E_i=\sqrt{k^2+m_i^2}}$ the energies of the on-shell propagating quasiparticles, $m_i$ their effective masses and
\begin{equation}
 \label{e:distfun}
  f_i^0 = (\exp(E_i/T)+S_i)^{-1} 
\end{equation}
the thermal equilibrium distribution functions with $S_{l,s}=1$ for fermions
and $S_g=-1$ for bosons. The degeneracy factors read explicitly $d_l=d_{\bar{l}}=2N_cN_l=12$
for $N_l=2$ light (anti-)quark flavours, $d_s=d_{\bar{s}}=2N_c=6$ for strange (anti-)quarks and
$d_g=2(N_c^2-1)=16$ for left- plus right-handed transversal gluons. The
effective quasiparticle masses depend on the dynamically generated self-energies
$\Pi_i$ via
\begin{equation}
 \label{equ:effmass}
  m_i^2=m_{i,0}^2+\Pi_i \,,
\end{equation}
where we include the current masses $m_{i,0}$ with $m_{g,0}=0$, $m_{l,0}=5~\textrm{MeV}$
and $m_{s,0}=95~\textrm{MeV}$. For $\Pi_i$ we use the asymptotic forms of the gauge
independent hard thermal loop self-energies~\cite{Bluhm:2006yh,Pisarski:1989wb}
\begin{eqnarray}
 \label{e:pig}
  \Pi_g(T) & = & \left(3+\frac{N_f}{2}\right)\frac{G(T)^2}{6}T^2 , \\
 \label{e:piq}
  \Pi_l(T) & = & 2 \left(m_{l,0} \sqrt{\frac{G(T)^2}{6}T^2}+
  \frac{G(T)^2}{6}T^2\right) , \\
 \label{e:pis}
  \Pi_s(T) & = & 2 \left(m_{s,0} \sqrt{\frac{G(T)^2}{6} T^2}+
  \frac{G(T)^2}{6} T^2\right) ,
\end{eqnarray}
where the perturbative running coupling has been replaced by an effective
coupling $G(T)$ which in the high-temperature regime resembles the perturbative coupling
for thermal momenta. This set-up of the model can be modified to describe pure
Yang-Mills thermodynamics by setting $N_f=d_l=d_{\bar{l}}=d_s=d_{\bar{s}}=0$.

\begin{figure}[t]
\centering
  \includegraphics[width=0.98\linewidth]{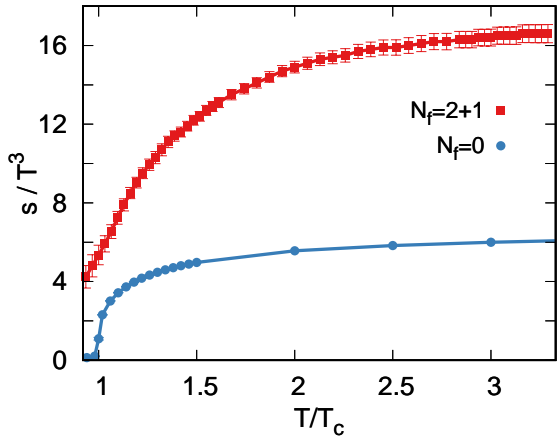}
\caption{(Colour online) Scaled entropy density $s/T^3$ as a function of scaled temperature $T/T_c$. The quasiparticle model results (solid lines) are shown in comparison to the lattice gauge theory results for Yang-Mills theory ($N_f=0$) from~\cite{Borsanyi:2012ve} (full blue circles) and for $N_f=2+1$ QCD from~\cite{Borsanyi:2013bia} (full red squares). We use $T_c=260~\textrm{MeV}$ and $T_c=155~\textrm{MeV}$ for $N_f=0$ and $N_f=2+1$, respectively.}
\label{fig:QPMBasics}
\end{figure}
\begin{figure*}[t]
\centering
  \includegraphics[width=0.48\linewidth]{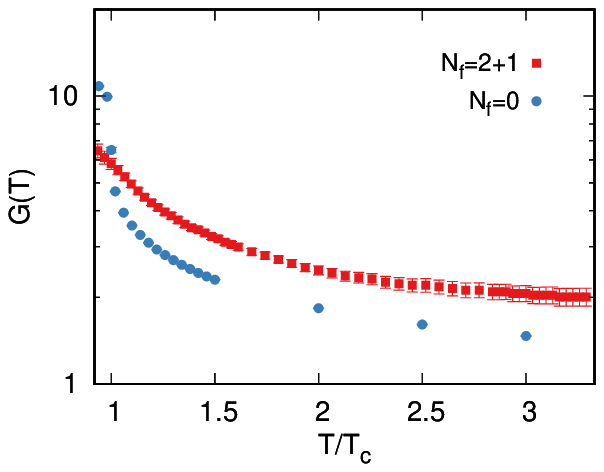}
  \includegraphics[width=0.48\linewidth]{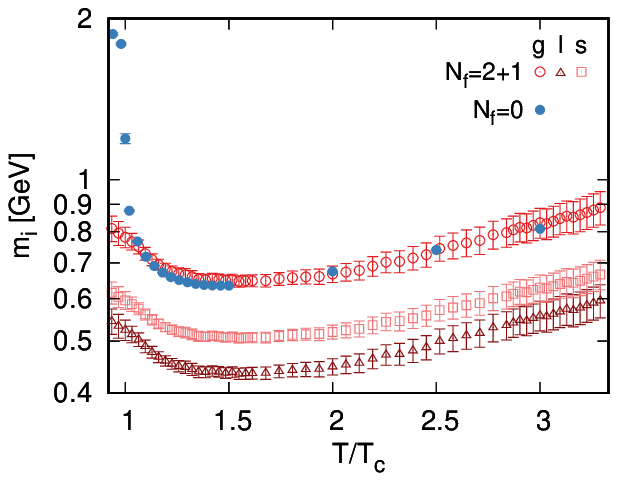}
  \caption{(Colour online) Left: Effective coupling $G(T)$ as a function of scaled temperature $T/T_c$ employed in the description of the scaled entropy density shown in Fig.~\ref{fig:QPMBasics}. Right: Corresponding effective quasiparticle masses $m_i(T)$, see Eq.~\eqref{equ:effmass}, (full blue circles and open red circles for gluons, open red triangles for light and open red squares for strange quarks) as functions of $T/T_c$. The error bars shown in both panels highlight estimates for the uncertainties obtained as a result of the errors in the lattice data for $s/T^3$.}
\label{fig:QPMInput}
\end{figure*}

Figure~\ref{fig:QPMBasics} shows results for the scaled entropy density $s/T^3$ in the quasiparticle model compared to state-of-the-art lattice gauge theory results for pure Yang-Mills theory~\cite{Borsanyi:2012ve} (circles) and $N_f=2+1$ QCD with physical quark masses~\cite{Borsanyi:2013bia} (squares). The temperature-dependence of the effective coupling $G(T)$ in the model is adjusted as to describe the lattice data and accommodate non-perturbative effects near the deconfinement transition temperature $T_c$. The results for $G(T)$ are shown in Fig.~\ref{fig:QPMInput} (left panel). The depicted error bars reflect possible variations in $G(T)$ as a consequence of the errors reported for the lattice data seen in Fig.~\ref{fig:QPMBasics}. The corresponding effective quasiparticle masses are shown in the right panel of Fig.~\ref{fig:QPMInput}.

While the entropy density in Yang-Mills theory exhibits indications for a first-order phase transition, see Fig.~\ref{fig:QPMBasics}, $s/T^3$ is continuous for $T$ around $T_c$ in $N_f=2+1$ QCD. Through the presence of dynamical quark degrees of freedom the scaled entropy density is increased by about a factor \mbox{$2-3$} in the deconfined phase. This is reflected in the behaviour of the effective coupling $G(T)$, see Fig.~\ref{fig:QPMInput} (left). Except for $T\lesssim T_c$, where the effective coupling and, thus, the gluon quasiparticle mass must become large in order to describe the sudden drop in the Yang-Mills entropy density, $G(T)$ is larger for QCD than for Yang-Mills theory. Moreover, at larger $T$ both couplings exhibit comparable slopes. This agrees with the perturbatively expected behaviour of the \mbox{$\beta$-function} and its $N_f$-dependence. With the corresponding temperature dependence of the effective quasiparticle masses, shown in Fig.~\ref{fig:QPMInput} (right), the QPM is capable of describing the lattice data for $s/T^3$. While $m_i/T$ at high $T$ vanishes logarithmically in line with the perturbative coupling for $k\sim T$, $m_i(T)$ itself rises approximately linearly with $T$ in this regime, exhibits a
minimum somewhat above $T_c$ and becomes large near $T_c$. Moreover, the gluon effective mass is found to be comparable for Yang-Mills theory and QCD when plotted as a function of $T/T_c$. The apparent $N_f$-independence in the shown temperature interval is a consequence of the compensation of two effects, the $N_f$-dependence in the dynamically generated gluon self-energy $\Pi_g(T)$ in Eq.~\eqref{e:pig} including the behaviour of $G(T)$ and the $N_f$-dependence of $T_c$.

\section{Kinetic theory in relaxation time approximation\label{Sec:RTA}}

In this work, we determine the shear viscosity of deconfined strongly interacting matter by making use of the Boltzmann kinetic transport equation which for each quasiparticle species $i$ with medium dependent dispersion relation reads as
\begin{equation}
\label{e:BoltzmannVlasov}
 \left(k_i^\mu \partial_\mu + m_i F_i^\mu \partial_{k_i^\mu} \right)f_i = {\cal C}_i[\{f_j\}] \,.
\end{equation}
The second term on the left-hand side of Eq.~\eqref{e:BoltzmannVlasov} contains an external force on the quasiparticles, $F_i^\mu=\partial^\mu m_i$ with $k_{i,\mu}F_i^\mu=0$, induced by the residual mean field interaction as a consequence of the temperature-dependent effective mass $m_i(T)$. This is essential when making contact between the kinetic theory description and fluid dynamics by defining a covariantly conserved energy-momentum tensor from which transport coefficients can be  determined~\cite{Dusling:2011fd,Bluhm:2012km}.

In the following we will consider the case of out of but near local thermal equilibrium. This allows us to expand the Boltzmann equation around its local thermal equilibrium solution $f_i^0$ such that the left-hand side of Eq.~\eqref{e:BoltzmannVlasov} can be written in terms of gradients of the thermodynamic variables and the collision operator ${\cal C}_i$, which formally depends on all $f_j$, becomes linearized in the deviation $\delta f_i = f_i-f_i^0$ from equilibrium. Furthermore, we will study the collision operator in relaxation time (or Bhatnagar-Gross-Krook) approximation which amounts to replacing~\cite{Reif}
\begin{equation}
\label{e:RTA}
 {\cal C}_i[\{f_j\}] = - \frac{k_i^\mu u_\mu}{\tau_i} \delta f_i \,,
\end{equation}
where $\tau_i$ is the energy-averaged relaxation time for species $i$ in the presence of other quasiparticles and $u_\mu$ is the fluid four-velocity field. In the local rest frame of the fluid we have $u_\mu=(1,\vec{0})$ and $k_i^\mu=(E_i,\vec{k})$.

In this approximation the leading-order deviation of the covariantly conserved energy-momentum tensor from local thermal equilibrium can easily be obtained by the sum of individual quasiparticle contributions. Matching the expression with its corresponding definition in fluid dynamics allows one to find an explicit form of the shear viscosity in the local rest frame which depends on the $\tau_i$. For a given quasiparticle species $i$ we \mbox{have~\cite{Hosoya:1983xm,Gavin:1985ph,Sasaki:2008fg,Bluhm:2009ef,Chakraborty:2010fr,Dusling:2011fd}}
\begin{equation}
\label{eta}
 \eta_i = \frac{1}{15 T} \int \frac{d^3k}{(2\pi)^3} \frac{\vec{k}^4}{E_i^2}
          d_i \tau_i f_i^0 (1 - S_i f_i^0) \,.
\end{equation}
For QCD with $N_f=2+1$ quark flavours at $\mu_B=0$ one finds, therefore, for the total shear viscosity
\begin{equation}
\label{etatot}
 \eta = \sum_i \eta_i = 2(\eta_{l} + \eta_{s}) + \eta_g \,,
\end{equation}
while in pure Yang-Mills theory we have $\eta=\eta_g$.

The essential quantities that need to be evaluated are the relaxation times $\tau_i$ entering Eq.~\eqref{eta}. In this work they are explicitly computed from the microscopic scattering cross sections for scatterings among massive quasiparticle excitations. The relaxation time is inversely related to the particle number density of scattering partners and the scattering cross section. For a multi-component system it follows in matrix form as~\cite{Reif} $\hat{\tau}^{-1}=\hat{n}\hat{\bar{\sigma}}$. For QCD with $N_f=2+1$ quark flavours this explicitly reads
\begin{equation}
\begin{pmatrix}
\tau^{-1}_l \\
\tau^{-1}_{\bar{l}} \\
\tau^{-1}_s \\
\tau^{-1}_{\bar{s}} \\
\tau^{-1}_g
\end{pmatrix}
=
\begin{pmatrix}
\bar{\sigma}_{ll} & \bar{\sigma}_{l\bar{l}} & \bar{\sigma}_{ls} & \bar{\sigma}_{l\bar{s}} & \bar{\sigma}_{lg} \\
\bar{\sigma}_{\bar{l}l} & \bar{\sigma}_{\bar{l}\bar{l}} & \bar{\sigma}_{\bar{l}s} & \bar{\sigma}_{\bar{l}\bar{s}} & \bar{\sigma}_{\bar{l}g} \\
\bar{\sigma}_{sl} &\bar{\sigma}_{s\bar{l}} & \bar{\sigma}_{ss} & \bar{\sigma}_{s\bar{s}} & \bar{\sigma}_{sg} \\
\bar{\sigma}_{\bar{s}l} & \bar{\sigma}_{\bar{s}\bar{l}} & \bar{\sigma}_{\bar{s}s} & \bar{\sigma}_{\bar{s}\bar{s}} & \bar{\sigma}_{\bar{s}g} \\
\bar{\sigma}_{gl} & \bar{\sigma}_{g\bar{l}} & \bar{\sigma}_{gs} & \bar{\sigma}_{g\bar{s}} & \bar{\sigma}_{gg}
\end{pmatrix}
\begin{pmatrix}
d_l\, n_l \\
d_{\bar{l}}\, n_{\bar{l}} \\
d_s\, n_s \\
d_{\bar{s}}\, n_{\bar{s}} \\
d_g\, n_g
\end{pmatrix}\,,
 \label{matrix}
\end{equation}
where
\begin{equation}
 n_{i} = \int \frac{d^3k}{(2\pi)^3}f_i^0
\end{equation}
is the $T$-dependent particle number density per degree of freedom of quasiparticle species $i$. While in general $\tau_i$ depends on the energy $E_i$ of the quasiparticle, we approximate $\tau_i$ by its energy-average which depends on the energy-averaged cross section $\bar{\sigma}_{ij}$. From Eq.~\eqref{matrix} one readily finds the relaxation time for light quarks as
\begin{eqnarray}
\tau^{-1}_{l} & = &
\frac{d_l}{2} n_{l} [\bar{\sigma}_{ud \rightarrow ud} + \bar{\sigma}_{uu \rightarrow uu}] + d_s n_s \bar{\sigma}_{us \rightarrow us}
\nonumber\\
& & + \,\frac{d_{\bar{l}}}{2} n_{\bar{l}} [\bar{\sigma}_{u\bar{u} \rightarrow u \bar{u}} + \bar{\sigma}_{u\bar{u} \rightarrow d \bar{d}} + \bar{\sigma}_{u\bar{u} \rightarrow s \bar{s}} + \bar{\sigma}_{u\bar{u} \rightarrow g g}
\nonumber\\
& & + \,\bar{\sigma}_{u\bar{d} \rightarrow u \bar{d}}]
\nonumber\\
& & + \,d_{\bar{s}} n_{\bar{s}} \bar{\sigma}_{u \bar{s} \rightarrow u \bar{s}} + d_g n_g \bar{\sigma}_{ug \rightarrow u g}
\label{taul}
\end{eqnarray}
or for gluons as
\begin{eqnarray}
\tau^{-1}_{g} & = &
d_g n_g [\bar{\sigma}_{gg \rightarrow gg} + \bar{\sigma}_{gg \rightarrow u\bar{u}} + \bar{\sigma}_{gg \rightarrow d\bar{d}} + \bar{\sigma}_{gg \rightarrow s\bar{s}}]
\nonumber\\
& & + \,d_l n_{l} \bar{\sigma}_{gu \rightarrow gu} + d_{\bar{l}} n_{\bar{l}} \bar{\sigma}_{g\bar{u} \rightarrow g\bar{u}}
\nonumber\\
& & + \,d_s n_s \bar{\sigma}_{gs \rightarrow gs} + d_{\bar{s}} n_{\bar{s}} \bar{\sigma}_{g\bar{s} \rightarrow g\bar{s}}
\label{taug}
\end{eqnarray}
while for pure Yang-Mills theory we have only \mbox{$\tau^{-1}_{g}=d_g n_g \bar{\sigma}_{gg \rightarrow gg}$.}

The individual energy-averaged cross sections for the scattering process $(1,2)\to(3,4)$ in the medium are given by~\cite{Zhuang:1995uf}
\begin{eqnarray}
\nonumber
 & & \overline{\sigma}_{12\to 34} (T) = \int_{s_{\rm th}}^\infty ds~ \int_{t_{\rm min}}^{t_{\rm max}} dt \,\frac{d\sigma_{12\to 34}}{dt}(s,t;T) \\
\label{sigmabar}
 & & \times \sin^2\theta(s,t;T) (1 - S_3 f_3^0) (1 - S_4 f_4^0) P(s;T) \,. \label{eff_cross}
\end{eqnarray}
We note that $\bar{\sigma}$ depends on $T$ both explicitly via the equilibrium distribution functions $f_i^0(s;T)$ and implicitly via $G(T)$ and $m_{i=1\dots 4}(T)$. In writing Eq.~\eqref{sigmabar} we have assumed that the center-of-mass (c.m.) of the system is at rest in the medium such that all entering quantities can be expressed in terms of the Mandelstam variables $s$ and $t$, where $u$ can be replaced using the condition $s+t+u=\sum_{i=1}^4 m_i^2$. Accounting for the possible phase space occupation in the final state, the factors $(1-S_if_i^0)$ represent Pauli blocking (for quarks and anti-quarks) or Bose enhancement (for gluons) in the medium. The integration limits in the four-momentum transfer $t_{\rm min}$ and $t_{\rm max}$ are determined from the condition $-1\leq\cos\theta\leq 1$, where $\theta$ is the scattering angle, while $s_{\rm th}={\rm max}[(m_1+m_2)^2,(m_3+m_4)^2]$. Moreover, as in~\cite{Zhuang:1995uf,Sasaki:2008um,Danielewicz:1984ww} we include the phenomenological weight-factor $\sin^2\theta$ in Eq.~\eqref{sigmabar} which signals the dominance of large angle scatterings for the transport of momentum. As a consequence, $\bar{\sigma}$ is reduced compared to the isotropic cross section which implies an increase in the $\tau_i$. Finally, $P(s;T)$ denotes the probability of finding $(3,4)$ with c.m. energy $s$ in the final state,
\begin{eqnarray}
\nonumber
 P(s;T) & = & C \sqrt{(s-m_1^2 - m_2^2)^2-4m_1^2 m_2^2} \\
 & & \times f_3^0 f_4^0 v_{\rm rel}(s;T) \,,
\end{eqnarray}
where the normalization constant $C$ is fixed via
\begin{equation}
 \int_{s_{\rm th}}^\infty ds\ P(s;T) = 1
\end{equation}
and $v_{\rm rel}(s;T)$ is the relative velocity between the two scattering quasiparticles
\begin{equation}
\label{vrel}
 v_{\rm rel}(s;T) = \frac{2s \sqrt{(s-m_1^2 -m_2^2)^2- 4 m_1^2 m_2^2}}{s^2-(m_1^2-m_2^2)^2} \,.
\end{equation}

The differential cross section $d\sigma/dt$ for the process $(1,2)\to(3,4)$ entering Eq.~\eqref{sigmabar} is obtained from the corresponding total scattering amplitude squared $\langle |\mathcal{M}|^2 \rangle$ via
\begin{eqnarray}
\nonumber
& & \frac{d\sigma_{12\to 34}}{dt} (s,t;T) = \frac{1}{16 \pi ((s-m_{1}^2-m_{2}^2)^2 - 4m_{1}^2 m_{2}^2)} \\
\label{differential}
& & \times \langle |\mathcal{M}_{12\to 34}|^2 (s,t;T) \rangle \,.
\end{eqnarray}
In $\langle |\mathcal{M}|^2 \rangle$ we sum over the spin/polarization and colour degrees of freedom in the final state and, since the degeneracy factors $d_i$ are included already in Eqs.~\eqref{eta} and~\eqref{matrix}, average over the initial state. The individual amplitudes are computed perturbatively at tree level for the elementary two-body scattering processes $qq \rightarrow qq$, $qq' \rightarrow qq'$, $q\bar{q} \rightarrow q\bar{q}$,  $q\bar{q}' \rightarrow q\bar{q}'$, $\bar{q}\bar{q} \rightarrow \bar{q}\bar{q}$, $gg \rightarrow gg$, $q\bar{q} \rightarrow q'\bar{q}'$, $q\bar{q} \rightarrow gg$ and $gg \rightarrow q\bar{q}$ among the massive quasiparticles, where $q=u,d,s$ and also exchanged gluons obey Eq.~\eqref{equ:effmass}. Accordingly, the (anti-)quark and gluon propagators are, suppressing colour-indices, modified as
\begin{equation}
\label{prop}
 \frac{i}{\Slash{k} - m_{l,\bar{l},s,\bar{s}}} \,,
 \quad
 \frac{-ig^{\mu\nu}}{k^2 - m_g^2} \,,
\end{equation}
respectively. Expressing the gluon propagator in Feynman gauge allows us to enforce directly the on-shellness condition for the quasiparticles in the thermal medium. For the coupling we employ the effective coupling $G(T)$. Explicit expressions for the scattering amplitudes will be reported elsewhere. We note, however, that in the limit $m_{i=1\dots 4}\to 0$ our analytic expressions for the differential cross sections agree with those presented in~\cite{Berrehrah:2013mua} and found in~\cite{Peskin:1995ev}.

\begin{figure}[t]
\centering
 \includegraphics[width=0.98\linewidth]{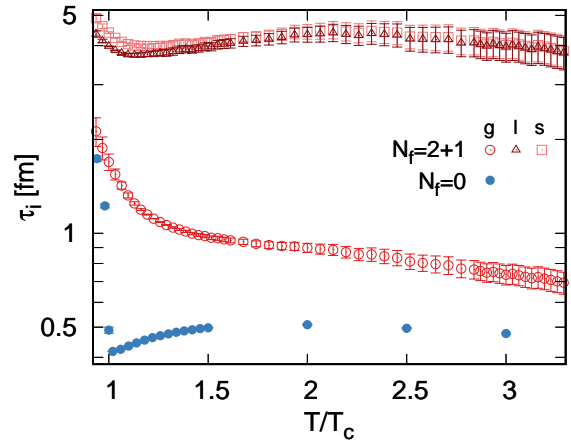}
 \caption{(Colour online) Relaxation times $\tau_i$ as functions of scaled temperature $T/T_c$ for different quasiparticle species. Pure Yang-Mills theory results for $\tau_g$ (full blue circles) are contrasted with $N_f=2+1$ quark flavour QCD results (open circles for a gluon, open triangles for a light and open squares for a strange quark). The QCD results shown for $\tau_l$ and $\tau_g$ are obtained via Eqs.~(\ref{taul}) and~(\ref{taug}), respectively.}
\label{fig:tau}
\end{figure}

With the above-described set-up we compute the relaxation times $\tau_i$ in pure Yang-Mills theory and for $N_f=2+1$ QCD. The corresponding results as functions of $T/T_c$ are presented in Fig.~\ref{fig:tau}. In pure Yang-Mills theory $\tau_g$ exhibits a sharp minimum around $T_c$ and a shallow maximum for about $2T_c$ before slowly decreasing with increasing temperature. The pronounced non-monotonicity near $T_c$ is caused by the behaviour of $G(T)$, see Fig.~\ref{fig:QPMInput} (left panel). A qualitatively similar observation can be made for light and strange quarks in QCD.

However, $\tau_l$ and $\tau_s$ are an order of magnitude larger, and both extrema are smooth and shifted towards slightly higher temperatures. Moreover, one observes that the current quark mass $m_{i,0}$ plays a considerable role only for $T<1.5\,T_c$. In contrast, $\tau_g$ in QCD remains a monotonically decreasing function of $T/T_c$ that is roughly a factor $4-5$ smaller than $\tau_{l,s}$. Since the shear viscosity directly depends on the relaxation time, see Eq.~(\ref{eta}), it is clear that the main contribution to the total shear viscosity in QCD will stem from the quark and anti-quark sectors. Furthermore, the increase of $\tau_g$ from pure Yang-Mills theory to QCD highlights the impact of dynamical quarks in the QGP on the effectiveness of gluons at equilibrating momentum degradations.

\section{Flavour dependence of the specific shear viscosity\label{Sec:Results}}

With the relaxation times $\tau_i$ at hand, we can now calculate the shear viscosity for pure Yang-Mills theory and $N_f=2+1$ QCD and compare systematically both theories to study the impact of the quark matter sector in QCD.

\begin{figure}[t]
\centering
  \includegraphics[width=0.98\linewidth]{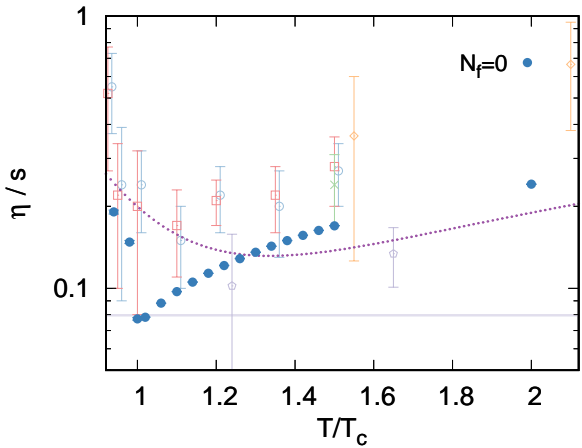}
\caption{(Colour online) Shear viscosity to entropy density ratio as a function of $T/T_c$ for pure Yang-Mills theory in the quasiparticle model (full blue circles). For comparison, the corresponding lattice gauge theory results from~\cite{Nakamura:2004sy} (open yellow diamonds), from~\cite{Meyer:2007ic} (open grey pentagons), from~\cite{Mages:2015rea} (green cross) and from~\cite{Astrakhantsev:2017nrs} (open blue circles and open red squares) are shown. The KSS-bound of $1/4\pi$ is indicated by the horizontal line and the parametric representation of the results from the functional diagrammatic approach~\cite{Christiansen:2014ypa} is shown by the dotted purple line.}
\label{fig:YMspecificshear}
\end{figure}
In Fig.~\ref{fig:YMspecificshear} we show first the temperature dependence of the shear viscosity to entropy density ratio for pure Yang-Mills theory (full blue circles). The ratio exhibits an abrupt, non-monotonic change in its behaviour around the first-order phase transition with a pronounced minimum at $T_c$ and a mild, monotonic increase for larger $T$. This behaviour can be traced back to the effective coupling $G(T)$ and the entropy density $s(T)$. It is an intriguing observation that the minimum of the specific shear viscosity reaches the KSS lower bound of $1/4\pi$. In Fig.~\ref{fig:YMspecificshear} we also compare our results with available data from lattice gauge theory calculations~\cite{Nakamura:2004sy,Meyer:2007ic,Mages:2015rea,Astrakhantsev:2017nrs} and with the results from employing the gluon spectral function in the functional diagrammatic approach~\cite{Christiansen:2014ypa}. Overall, our results agree remarkably with the bulk of information from first-principles. The global behaviour found in~\cite{Christiansen:2014ypa} (see dotted purple line in Fig.~\ref{fig:YMspecificshear} for a parametric representation) is within the reported errors well captured by our model in a wide range of temperatures, in particular for $T$ above $1.3\,T_c$. However, near $T_c$ we find a significantly stronger non-monotonicity with a minimal $\eta/s$ around $T_c$ instead of slightly above $T_c$ as a natural consequence of the first-order phase transition.

\begin{figure}[t]
\centering
  \includegraphics[width=0.98\linewidth]{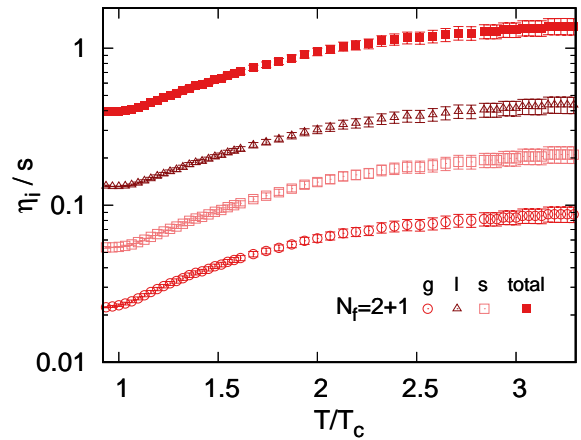}
\caption{(Colour online) Shear viscosity to entropy density ratio as a function of $T/T_c$ in $N_f=2+1$ QCD. The individual contributions $\eta_i/s$ from gluons (circles), light quarks (triangles, as sum of up and down quarks) and strange quarks (open squares) with equal contributions from their anti-quarks are shown, along with the total specific shear viscosity of the QGP (full squares), corresponding to Eq.~(\ref{etatot}).}
\label{fig:eta}
\end{figure}
The ratio $\eta/s$ in QCD with $N_f=2+1$ quark flavours is exhibited in Fig.~\ref{fig:eta} (full squares), where the individual contributions from the light (as the sum of up and down quark contributions) and strange quark sectors as well as from gluons to the total $\eta/s$ are also presented. We find a rather shallow minimum of about $0.4$ around the pseudo-critical temperature $T_c$ and a moderate, monotonic increase with $T$ at larger temperatures for the total ratio. Similar behaviour can be seen for the individual contributions $\eta_i/s$ entering Eq.~\eqref{etatot}. This is a consequence of the dynamics encoded in the quasiparticle masses via the effective coupling $G(T)$. Moreover, one clearly observes a hierarchy among the individual contributions that follows inversely the ordering in the effective quasiparticle masses. As expected, the heaviest quasiparticles are the most effective ones in equilibrating momentum degradations within the QGP. We note that a similar but quantitatively different pattern was reported in~\cite{Chandra:2012qq}. We find $\eta_s/\eta_l<0.5$ approaching only slowly $0.5$ with increasing $T$ while $\eta_g/\eta_l\leq 0.2$ in the shown temperature-interval.

\begin{figure*}
\centering
  \includegraphics[width=0.48\linewidth]{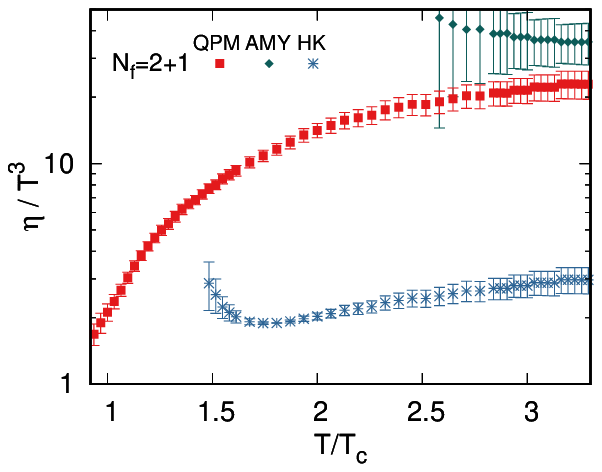}
  \includegraphics[width=0.48\linewidth]{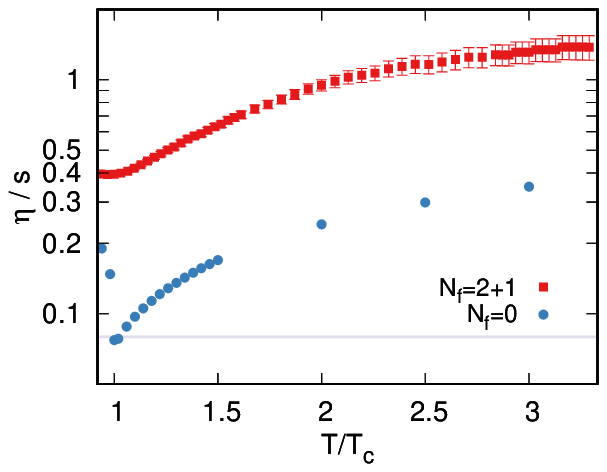}
    \caption{(Colour online) Left: Scaled shear viscosity $\eta/T^3$ as a function of scaled temperature $T/T_c$ for $N_f=2+1$ QCD in the quasiparticle model (QPM, red squares). For comparison, two different results for $N_f=3$ based on perturbative QCD calculations are shown: a) using the next-to-leading-log coupling-expansion result from Arnold, Moore and Yaffe~\cite{Arnold:2003zc} (AMY, green diamonds), and b) applying the parametrized relaxation time at leading-log order given by Hosoya and Kajantie~\cite{Hosoya:1983xm} (HK, blue stars). Right: Comparison of the specific shear viscosity as a function of $T/T_c$ between pure Yang-Mills theory (blue circles) and $N_f=2+1$ QCD (red squares) in the quasiparticle model.}
\label{fig:etaComparison}
\end{figure*}
In Fig.~\ref{fig:etaComparison} (left panel), we compare the shear viscosity of the QGP obtained in the present quasiparticle model with perturbative QCD expectations. The next-to-leading-log~(NLL) expansion in the running coupling $g$ as derived by Arnold, Moore and Yaffe~\cite{Arnold:2003zc} gives the following result for the shear viscosity
\begin{equation}
 \eta_{\textrm{NLL}} = \frac{T^3}{g^4} \left[\frac{\eta_1}{\ln(\mu_*/m_D)}\right]
 \label{AMY}
\end{equation}
with coefficients \mbox{$\eta_1=106.66$} and \mbox{$\mu_*/T=2.957$} for \mbox{$N_f=3$}, and Debye screening mass
\begin{equation}
 m_D^2 = \frac{1}{3} \left[C_A + N_f C_F \frac{d_F}{d_A} \right] g^2 T^2 \,,
\end{equation}
where $d_F=C_A=3$, $C_F=4/3$ and $d_A=8$. Another perturbative parametrization of the shear viscosity was proposed by Hosoya and Kajantie~\cite{Hosoya:1983xm}, reading
\begin{eqnarray}
 \eta & = & \frac{64 \pi^4}{675} \frac{{T^3}}{g^4 \ln(4\pi/g^2)} \left[ \frac{21\,N_f}{6.8\,[1+0.12(2 \,N_f+1)]}\right. \nonumber \\ & & + \left.\frac{16}{15\,[1+0.06\,N_f]} \right].
 \label{HK}
\end{eqnarray}
Here, we use $N_f=3$ for the number of quark flavours. The two terms in the square brackets of Eq.~\eqref{HK} mark contributions from massless quarks and gluons, respectively, which are both proportional to a relaxation time parametrized at leading-log order in $g$.

Replacing the running coupling in the perturbative expressions by our effective coupling $G(T)$, we find that the scaled shear viscosity, $\eta/T^3$, in the quasiparticle model approaches within errors the expectations from Eq.~(\ref{AMY}) at higher temperatures (see red squares and green diamonds in Fig.~\ref{fig:etaComparison} (left panel) for QPM and AMY, respectively). The difference between the QPM result and the result using Eq.~(\ref{HK}) is, however, significant (see blue stars in Fig.~\ref{fig:etaComparison} (left panel) for HK). This is a consequence of the fact that the latter describes a system of massless quarks and gluons. We can, thus, directly see the influence of the dynamical quasiparticle masses on the shear viscosity of the QGP. 
We note that applying the three approaches to pure Yang-Mills theory yields a similar result for the scaled shear viscosity.

The direct comparison between the quasiparticle model results of $\eta/s$ for $N_f=2+1$ QCD and pure Yang-Mills theory reveals a significant impact of the quark sector contributions in the entire range of temperatures studied in this work. This is shown in Fig.~\ref{fig:etaComparison} (right panel). The sizeable increase of $\eta/s$ in the presence of dynamical quarks is in line with the observations made for the relaxation times, see Fig.~\ref{fig:tau}. Although the entropy density is about a factor $2-3$ larger in QCD, this is not sufficient to balance the overall dominance of the quark and anti-quark contributions. Moreover, the pronounced non-monotonicity at $T_c$ in Yang-Mills theory is significantly smoothened in QCD, reflecting the difference in the order of the underlying phase transition.

Our results for $\eta/s$ in $N_f=2+1$ QCD are in quantitative contrast to the functional estimate for QCD reported in~\cite{Christiansen:2014ypa} which indicated only a moderate increase of the ratio for given $T/T_c$ due to the presence of dynamical quarks. While one might argue that our findings are within errors still compatible with the old lattice gauge theory results for pure Yang-Mills theory~\cite{Nakamura:2004sy} by Nakamura and Sakai, the bulk of available first-principle information is well overestimated by our QCD results. Moreover, we find a minimal $\eta/s$ that is at best at the very upper edge of possible values extracted for the QGP in early fluid dynamical applications. We note, however, that a very similar minimal value of the specific shear viscosity was found in other strongly coupled quantum fluids, namely ultra-cold atomic Fermi gases at or near the unitary limit. The shear viscosity of these physical systems can be studied experimentally, similar to flow experiments in heavy-ion collisions, through the fluid dynamical expansion of a trapped gas after removing the trapping potential~\cite{Cao:2010wa,Elliott:2013my,Joseph:2015my}. Analyzing these experiments with a proper fluid dynamical framework~\cite{Bluhm:2015raa} allows one to extract $\eta$ in the normal fluid phase as a function of temperature and density. In a recent study~\cite{Bluhm:2017rnf}, a minimal specific shear viscosity of $\eta/s=0.5\pm 0.1$ was found just above the transition temperature to superfluidity. Moreover, an increase of $\eta/s$ with $T$ in line with kinetic theory predictions could be extracted~\cite{Bluhm:2015bzi} supporting the idea of an underlying quasiparticle picture for the strongly coupled fluid.

\begin{figure}[t]
\centering
  \includegraphics[width=0.98\linewidth]{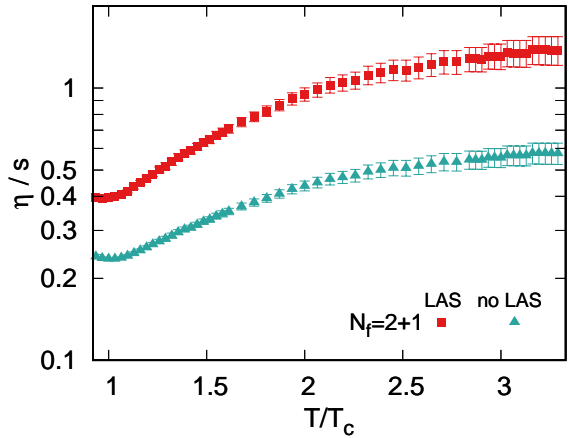}
   \caption{(Colour online) Specific shear viscosity as a function of $T/T_c$ for $N_f=2+1$ QCD in the quasiparticle model employing either the large angle scattering approximation (LAS, red squares) or the isotropic energy-averaged cross sections (no LAS, turquoise triangles).}
   \label{fig:las}
\end{figure}

Let us finally note that the quasiparticle model results presented in this work depend systematically on the approximations we made. One of these is the assumption of a large angle scattering dominance~\cite{Zhuang:1995uf,Sasaki:2008um,Danielewicz:1984ww} for the transport of momentum, see Eq.~\eqref{sigmabar}. Relaxing this approximation leads to an increase of the energy-averaged cross sections $\bar{\sigma}$ and, thus, to a reduction of the relaxation times and the specific shear viscosity. This is shown for $N_f=2+1$ QCD in Fig.~\ref{fig:las}, where we contrast $\eta/s$ computed with employing the large angle scattering approximation (LAS) and with using the full isotropic cross sections (no LAS). We note that only by employing the isotropic cross sections we find a temperature-dependence of the specific shear viscosity for the QGP that is quantitatively compatible with the results presented in the recent study~\cite{Moreau:2019vhw}.

\section{Summary\label{Sec:Conclusions}}

We have investigated the temperature dependence of the specific shear viscosity in pure Yang-Mills theory and in QCD with $N_f=2+1$ quark flavours in a quasiparticle model approach using kinetic theory in the relaxation time approximation. The effective, temperature-dependent masses in the quasiparticle dispersion relations are adjusted as to describe the equilibrium entropy density provided by first-principle lattice gauge theory simulations. Interestingly, we find that the gluon thermal mass for pure Yang-Mills theory and QCD is compatible when plotted as a function of scaled temperature $T/T_c$. This is a consequence of the compensation of the $N_f$-dependence in the deconfinement transition temperature $T_c$ and in the gluon self-energy. The relaxation times of the individual quasiparticle species are computed based on the microscopic scattering amplitudes of all the elementary two-body scatterings among the massive quasiparticles. For the associated coupling we employ the effective coupling of the model which enters the quasiparticle masses.

The shear viscosity to entropy density ratio, $\eta/s$, exhibits a sharp minimum at $T_c$ in pure Yang-Mills theory which coincides with the KSS bound $1/4\pi$ conjectured via gauge-gravity duality. The result near $T_c$ is found to be consistent with all the available lattice data within the errors. Moreover, the behavior at temperatures higher than $1.3\,T_c$ agrees fairly well with the functional diagrammatic approach~\cite{Christiansen:2014ypa}. Introducing quark-quasiparticles strongly modifies the temperature dependence of $\eta/s$ in QCD with $N_f=2+1$. The pronounced non-monotonic structure of the ratio at $T_c$ in pure Yang-Mills theory is replaced by a smooth behaviour with a shallow minimum around the pseudo-critical temperature in QCD. This modification is also reflected in the behaviour of the quasiparticle relaxation times.

In contrast to the functional estimate of $\eta/s$ for QCD reported in~\cite{Christiansen:2014ypa}, our microscopic calculations reveal a major impact of the dynamics carried by quark-quasiparticles as relevant effective degrees of freedom on top of the gluons. The non-trivial dynamics of those quasiparticles enters the scattering cross sections, which results in significant contributions to the specific shear viscosity. Another intriguing observation is that the quasiparticle approach yields a scaled shear viscosity, $\eta/T^3$, rather comparable in magnitude to the perturbative QCD result from the next-to-leading-log expansion~\cite{Arnold:2003zc} at a temperature of about $3\,T_c$. The comparison to the parameterized $\eta/T^3$ for massless quarks and gluons~\cite{Hosoya:1983xm}, on the other hand, exhibits a clear difference at any temperature studied in this work.

We have also illustrated the impact of the large angle scattering (LAS) approximation which was applied to
evaluate the energy-averaged cross sections. It is shown that the LAS prescription yields systematically larger contributions to $\eta/s$ than employing fully isotropic cross sections. With the latter prescription, we find quantitatively comparable results to those reported in \cite{Moreau:2019vhw}.

It is a straightforward application of what we have developed in this paper to study other transport coefficients and their phenomenological impact on observables via viscous fluid dynamical simulations. Also, the $\mu_B$-dependence of $\eta/s$ and other transport coefficients may be investigated. This will be reported elsewhere.

\acknowledgments
This work was partly supported by the Polish National Science Center (NCN), under Maestro grant number DEC-2013/10/A/ST2/00106 (K.R. and C.S.). The work of V.M. and M.B. is partially supported by the European Union's Horizon 2020 research and innovation program under the Marie Sk\l{}odowska Curie grant agreement No 665778 via the Polish National Science Center (NCN) under Polonez grant number UMO-2016/21/P/ST2/04035. M.B. also acknowledges the support by the program ``Etoiles montantes en Pays de la Loire 2017''. The authors thank B.~K\"ampfer, M.~Nahrgang and J.~Pawlowski for stimulating discussions. K.R. also acknowledges support of the Polish  Ministry of Science and Higher Education.


\end{document}